\documentstyle[12pt]{article}
\input epsf
\begin{document}
\vspace*{3cm}
\begin{center}
{\Large {\bf NON--GAUSSIAN FLUCTUATIONS FROM}}\\
\vspace*{0.2cm}
{\Large {\bf TEXTURES}}\\
\vspace*{1cm}
{\large Alejandro Gangui} \\
{\em ICTP -- International Center for Theoretical Physics,\\
P.~O.~Box 586, 34100 Trieste, Italy.}\\
and\\
{\em SISSA -- International School for Advanced Studies,\\
Via Beirut 4, 34013 Trieste, Italy.}

\vspace*{20pt}
{\bf Abstract}
\end{center}

\noindent
One of the most powerful tools to probe the existence of cosmic
defects in the early universe is through the 
Cosmic Microwave Background (CMB) radiation. 
It is well known that computations with causal sources 
are more involved than the adiabatic counterparts based on inflation, 
and this fact has in part 
hampered the development of fine detailed predictions. 
Analytical modeling, while necessarily limited in power, 
may tell us the overall characteristics of CMB from defects 
and hint at new features.
We apply an analytical model for textures to the study of 
non--Gaussian features of the CMB sky and  
compare our predictions with the four--year {\sl COBE}--DMR data.

\vspace*{40pt}
\begin{center}
To appear in Proceedings of the Moriond Conference on Microwave
Background Anisotropies, March 16th--23rd 1996 \\
\vspace*{20pt}
\end{center}
\newpage
\section{Introduction}
It has by now become clear that one of the most promising ways to learn
about the early universe is through the Cosmic Microwave Background
(CMB) radiation. 
With the prospective launch of future missions, 
like MAP and COBRAS/SAMBA\footnote{The best place to learn about these
missions and to follow the developments are, respectively,
the sites http://map.gsfc.nasa.gov/ and 
http://astro.estec.esa.nl/SA-general/Projects/Cobras/cobras.html}, 
one can hope that many of the so far elusive cosmological parameters
will be pinned down with unprecedent precision. 
The four--year {\sl COBE} plus other large--scale structure data 
have placed strong constraints on current models of structure formation. 
However the remaining window is still too large and many (widely) 
different cosmological models pass the test. 
This is actually the case with cosmic defect theories, on the one hand 
and inflation--based adiabatic models on the other.

The search for the so--called Doppler peaks (or Sakharov peaks if you like) 
seems to be among the first goals for next generation detectors, the aim being
in trying to discriminate among, say, standard adiabatic CDM models$^{1)}$,
cosmic strings$^{2)}$ and textures$^{3)}$. 
Whereas the generation of CMB anisotropies seems to be fully
understood within adiabatic models [refer to Hu's contribution to
these proceedings], the same does not happen for the latter models, 
where the non--linear evolution of the defects and their active role in
seeding anisotropies in the CMB makes the analysis far more involved
[see Durrer's contribution].
Moreover, causal sources can produce spectra mimicking the outcome of 
inflationary models$^{4)}$, hence increasing the
uncertainty and calling for fast and accurate methods for computing
the theoretical predictions [Seljak's contribution], 
and refined experiments to confront these competing theories. 

Other means of narrowing somewhat the window regards the recognition that 
the CMB radiation carries valuable information of the processes that
generated the anisotropies, all along the path of the photons from the
last scattering surface to our present detectors: 
should future experiments find (with high confidence level) a
departure of the statistics of the anisotropies from Gaussianity, 
it would disfavor the standard inflaton field quantum fluctuation origin 
of the cosmological perturbations. 
It then follows that it is interesting to calculate what predictions cosmic
defects make regarding non--Gaussian features in the CMB sky.
It is the aim of this short contribution to report on some work 
done on the CMB three--point correlation function predicted by textures
within a simple analytical model.

\section{The model}
Magueijo$^{5)}$ recently proposed a simple analytical model for the
computation of the ${\cal C}_\ell$'s from textures. The model exploits
the fact that in this scenario the microwave sky will show evidence of
spots
due to perturbations in the effective temperature of the photons
resulting from the non--linear dynamics of concentrations of
energy--gradients of the texture field. 
The model of course does not aim to replace the full range numerical
simulations but just to show overall features predicted by textures in
the CMB anisotropies.
In fact the model leaves free a couple of parameters that are fed in
from numerical simulations, like the number density
of spots, $\nu$, the scaling size, $d_s$, and
the brightness factor of the particular spot, $a_k$, telling
us about its temperature relative to the mean sky temperature.

Texture configurations giving rise to spots in the CMB are assumed to
arise with a constant probability per Hubble volume and Hubble time.
In an expanding universe one may compute the surface probability
density of spots
\begin{equation}
dP = N(y) dy d\Omega, \quad \mbox{with} \quad 
N(y) = - {8 \nu \ln (2) \over 3} \left(2^{y/3} - 1 \right)^2, 
\end{equation}
where $\Omega$ stands for a solid angle on the two--sphere and 
the time variable $y(t) \equiv \log_{2} (t_0 / t)$ measures how
many times the Hubble radius has doubled since proper time $t$ 
up to now\footnote{e.g., for a redshift $z_{\rm ls}\sim 1400$ 
at last scattering we have
$y_{\rm ls} \simeq \log_{2}[(1400)^{3/2}] \simeq 16$.}.

In the present context the anisotropies arise from 
the superposition of the
contribution coming from all the individual spots $S_k$ produced from
$y_{\rm ls}$ up to now, and so, 
$\Delta T / T  = \sum_{k} a_{k} S_k(\theta_k, y)$, where the random
variable $a_{k}$ stands for the brightness of the hot/cold $k$--th 
spot with characteristic values to be extracted 
from numerical simulations$^{6)}$. $S_k (\theta_k, y)$ is the
characteristic shape of the spots produced at time $y$, where
$\theta_k $ is the angle in the sky measured with respect to the
center of the spot.
A spot appearing at time $y$ has typically a size
$\theta^s(y) \simeq d_s  ~ \theta^{\rm hor}(y)$, with
$\theta^{\rm hor}(y)$ the angular size of the horizon at
$y$, and where it follows that 
$\theta^s(y) = \arcsin \left( {0.5 d_s   \over 2^{y/3} - 1} \right)$.
Textures are essentially causal seeds and therefore the spots induced
by their dynamics cannot exceed the size of the horizon
at the time of formation, hence $d_s \leq 1$.
Furthermore the scaling hypothesis implies that the profiles satisfy
$S_k(\theta_k , y) = S(\theta_k / \theta^s(y))$.
From all this it follows a useful expression for the multipole 
coefficients, $a_{\ell}^{m} =
\sum_k a_k S^\ell_k (y) {Y_{\ell}^{ m}}^* (\hat\gamma _k )$,
with $S^\ell_k (y)$ the Legendre transform of the spot profiles.

At this point the ${\cal C}_\ell$'s are easily calculated$^{5)}$.
As we are mainly concerned with the three--point function we go on and
compute the angular bispectrum predicted within this analytical model,
which we find to be  
\begin{equation}
\bigl\langle a_{\ell_1 }^{m_1} a_{\ell_2}^{ m_2} a_{\ell_3}^{ m_3}
\bigr\rangle =
\bigl\langle a^3 \bigr\rangle
\int dy N(y)  S^{\ell_1}(y) S^{\ell_2}(y) S^{\ell_3}(y)
\int d\Omega_{\hat\gamma} Y_{\ell_1}^{m_1} (\hat\gamma)
Y_{\ell_2}^{m_2}(\hat\gamma) Y_{\ell_3}^{m_3} (\hat\gamma) .
\end{equation}
$\bigl\langle  a^3 \bigr\rangle$ is the mean cubic value
of the spot brightness.

Having the expression for the bispectrum we may just plug it in the
formulae for the full mean three--point temperature correlation 
function$^{7)}$. To make contact with experiments however we 
restrict ourselves to the collapsed case where two out of the three
legs of the three--point function {\it collapse} and only one angle,  
say $\alpha$, survives (this is in fact one
of the cases analyzed for the four--year {\sl COBE}--DMR data$^{8)}$).

The collapsed three--point function thus calculated,
$\bigl\langle C_3 (\alpha )\bigr\rangle$,
corresponds to the mean value expected in an
ensemble of realizations. However, as we can observe just one particular
realization, we have to take into account the spread of the
distribution of the three--point function values when comparing
a model prediction with the observational results. 
This is the well--known cosmic variance problem. 
We can estimate the range of expected values about the mean by the
{\it rms} dispersion
$\sigma_{CV}^2(\alpha ) \equiv
\bigl\langle C_3^2 (\alpha ) \bigr\rangle - \bigl\langle C_3 (\alpha )
\bigr\rangle^2$.
We will estimate the range for the amplitude of the
three--point correlation function predicted by the model by
$\bigl\langle  C_3 (\alpha ) \bigr\rangle \pm \sigma_{CV}(\alpha )$.

It has been shown$^{6)}$ that spots generated from 
random field configurations of concentrations of
energy gradients lead to peak anisotropies 20 to 40 \% smaller than
those predicted by the spherically symmetric self--similar texture 
solution. These studies also suggest an asymmetry
between maxima $\langle a_{\rm max} \rangle$ and minima
$\langle a_{\rm min} \rangle$ of the peaks as being due to the fact
that, for unwinding events, the minima are
generated earlier in the evolution (photons climbing out of the
collapsing texture) than the maxima (photons falling in the collapsing
texture), and thus the field correlations are stronger for the maxima,
which enhance the anisotropies.

\section{Results}

Let us now compute the predictions on the CMB 
non--Gaussian features derived from the present analytical texture
model. 
One needs to have the distribution of the spot brightness
$\{a_k\}$ in order to compute the mean values $\langle a^n \rangle$.
It is enough for our present purposes to take for all hot spots 
the same $a_h > 0$ and for all the cold spots the same $a_c < 0$.
Then the $\langle a^n \rangle$
needed can readily be obtained in terms of $\langle a^2\rangle$ and
$x\equiv \langle a \rangle /\langle |a| \rangle$.
We fix $\langle a^2\rangle$ from the amplitude of the anisotropies
according to four--year {\sl COBE}--DMR$^{8)}$. 
The other
parameter, $x$, that measures the possible {\it asymmetry} between hot
and cold spots, we leave as a free parameter.

We consider the {\sl COBE}--DMR window function and, in order to take
into account the partial sky coverage due to the cut in the maps at Galactic
latitudes $\vert b \vert < 20^\circ$, we multiply $\sigma_{CV}$ by 
a factor $\sim 1.5$ in the numerical results 
(sample variance)\footnote{In the analysis of the data, 
the method used for computing the uncertainty is to generate
2000 random skies with HZ signal + noise, then to compute the 
three--point function of each realization 
{\it on the cut sky} after subtracting a best--fit multipole.
Hence, the result automatically includes sample variance.}.
\begin{figure}[t]
\centering
\leavevmode\epsfysize=10.5cm \epsfxsize=13.5cm \epsfbox{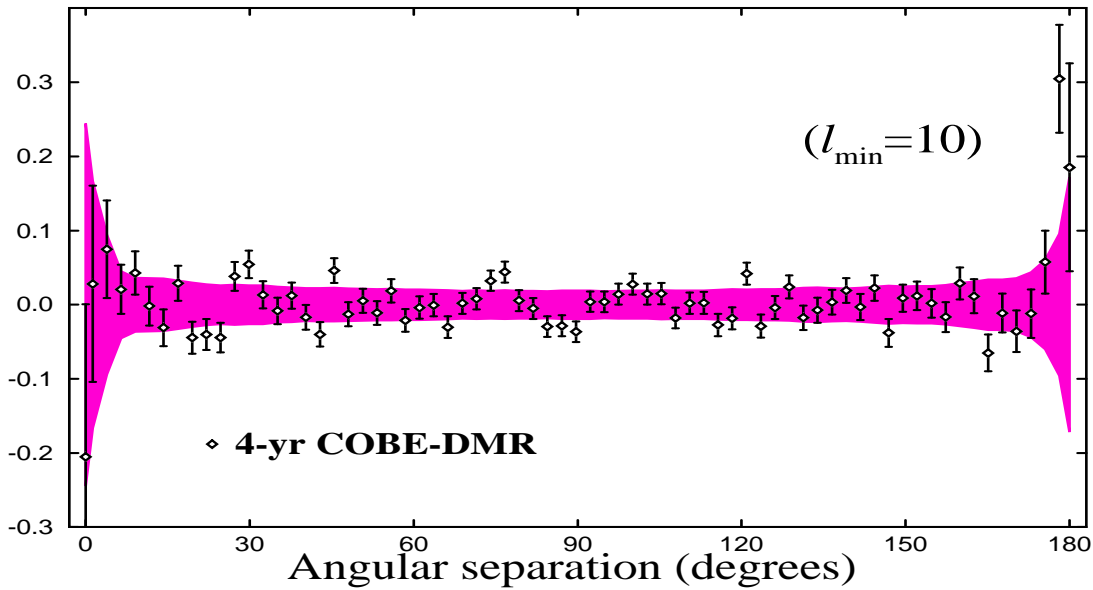}\\
\vspace{-0.2cm}
\parbox{16cm}{\caption[figure1]{\small
\baselineskip 12pt
The `pseudo--collapsed' three--point function
(in units of $10^4 {\mu}{\rm K}^3$)
as computed from the analysis of the four--year {\sl COBE}--DMR data
[data courtesy of G. Hinshaw and the {\sl COBE} team].
The points plotted are the analogue of our 
$\bigl\langle C_3 (\alpha ) \bigr\rangle$.
Error bars represent instrument noise while the grey band represents
the rms range of fluctuations due to a superposition of instrument
noise and cosmic variance.}}
\label{figuC-1}
\end{figure}
Let us now compare with the data: Subtracting the dipole and for all
reasonable values of the asymmetry parameter $x$, 
the data falls well within the 
$\bigl\langle  C_3 (\alpha ) \bigr\rangle \pm \sigma_{CV}(\alpha )$
band, and thus there is good agreement with the observations. 
However, the band for Gaussian distributed fluctuations
(e.g., as predicted by inflation) also 
encompasses the data well enough, and
it is in turn included inside the texture predicted band.
The fact that the range of
expected values for the three--point correlation function predicted by
inflation is included into that predicted by textures for all the
angles, and that the data points fall within them, makes it impossible
to draw conclusions favoring one of the models.

It is well known that the largest contribution to the cosmic variance
comes from the small values of $\ell$. 
Thus, no doubt the situation may improve if one subtracts
the lower order multipoles contribution, as in a 
$\ell_{min}=10$ analysis$^{9)}$.
In Figure 1 we show the analysis of the four--year {\sl COBE}--DMR
data evaluated from the 53 + 90 GHz combined map, containing power
from the $\ell = 10$ moment and up. 
It is apparent that the fluctuations about zero correlation 
(i.e., no signal) are too large for the instrument noise to be the
only responsible. These are however consistent with the range of
fluctuations expected from a Gaussian process (grey band). 

What we want to see now is whether our analytical texture model 
for the three--point function$^{10)}$ can do better when compared 
with the data. Figure 2 shows the collapsed three--point function
$\bigl\langle C_3 (\alpha ) \bigr\rangle$  (solid line) and the grey
band indicates the {\it rms} range of fluctuations expected from the
cosmic variance.
Also shown is a black band with the expected range for inflationary 
models (no instrument noise included).
The bands do not superpose each
other for some ranges of values of the separation angle for the value
of $x=0.4$ considered, what means that measurements in that ranges can
distinguish among the models. 
The value of the parameter $x$ considered is quite in excess of that 
suggested by simulations$^{6)}$, 
but was chosen to show an example with a noticeable effect.

\begin{figure}[t]
\centering
\leavevmode
{\hbox %
{\epsfxsize = 8.5cm \epsffile{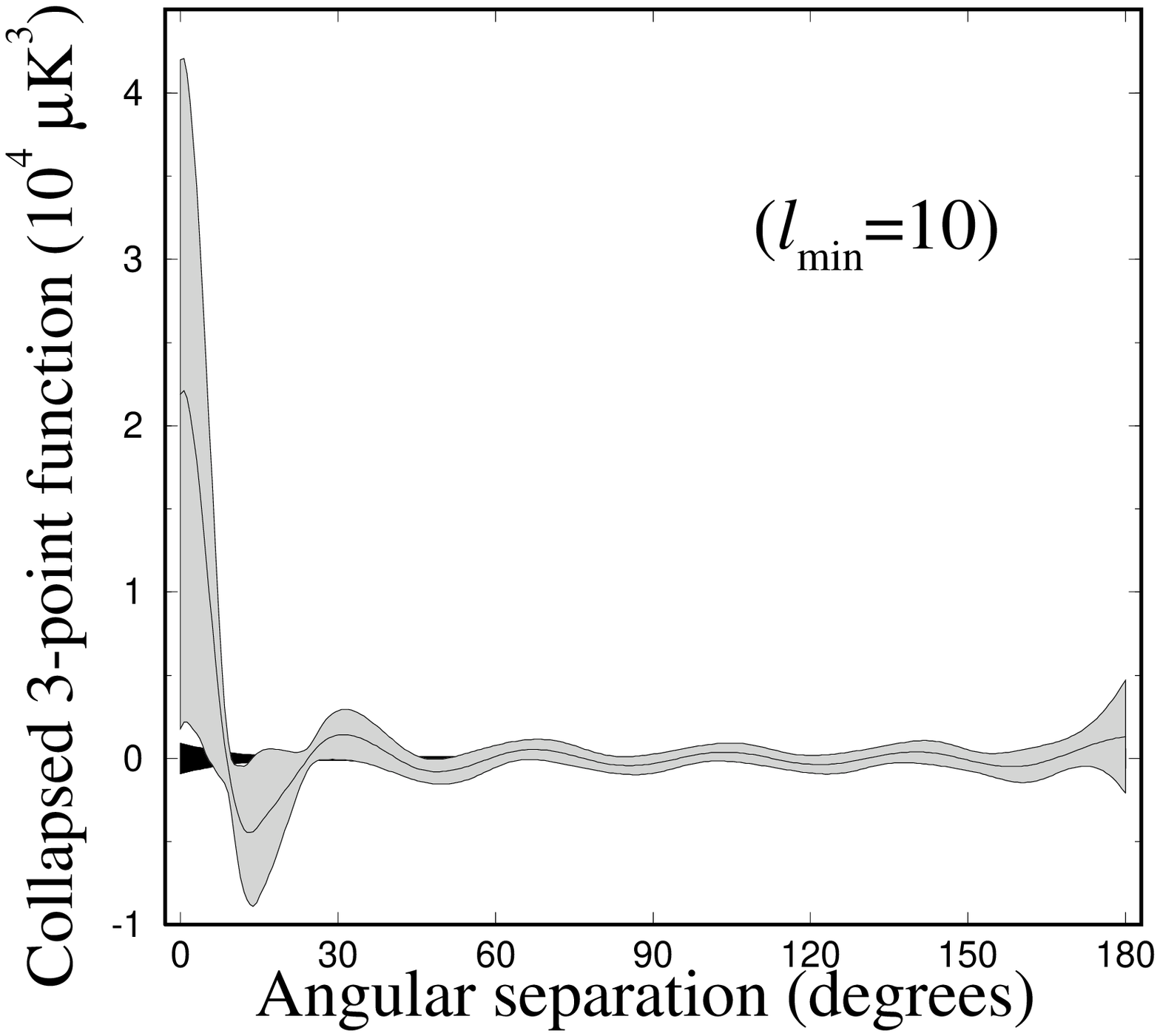} }
{\hskip -0.3cm} 
{\epsfxsize = 8.5cm \epsffile{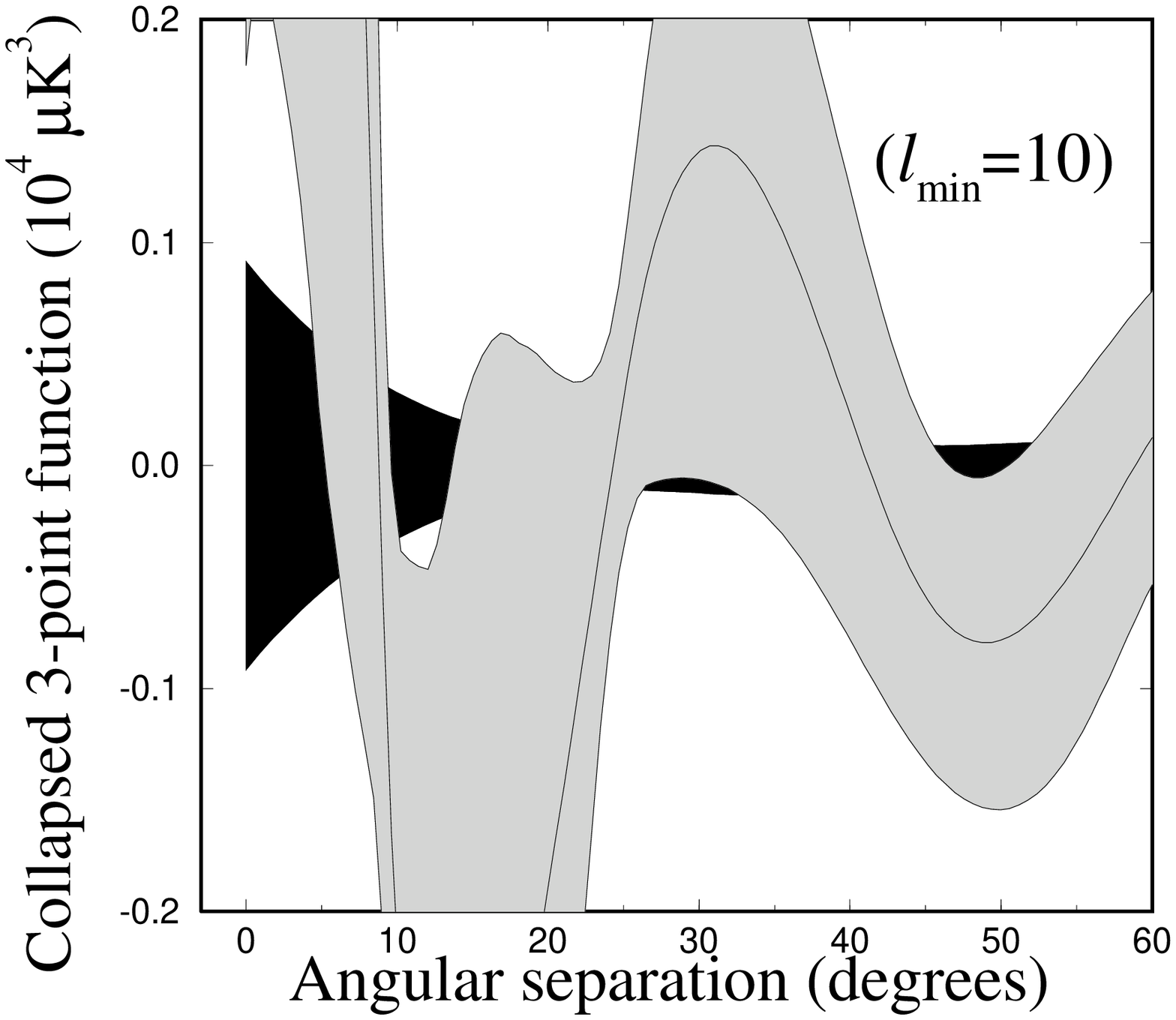} } }
\parbox{16cm}{\caption[figure2]{\small
\baselineskip 12pt
Expected range of values for the three--point correlation
function for the texture model
with asymmetry parameter $x=0.4$
(grey band).
Also shown is the expected range for inflationary models (black band).
Both bands include the $\sim 50\%$ increase in $\sigma_{CV}$ due to
the sample variance.
All multipoles up to $\ell = 9$ have been subtracted.
The right panel
shows a zoomed fraction of the same plot.}}
\end{figure}

In Figure 3 we show the result of combining the previous figures,
confronting the actual data with the curves predicted by the texture
model. 
\begin{figure}[t]
\centering
\leavevmode
{\hbox %
{\epsfxsize = 8.5cm \epsffile{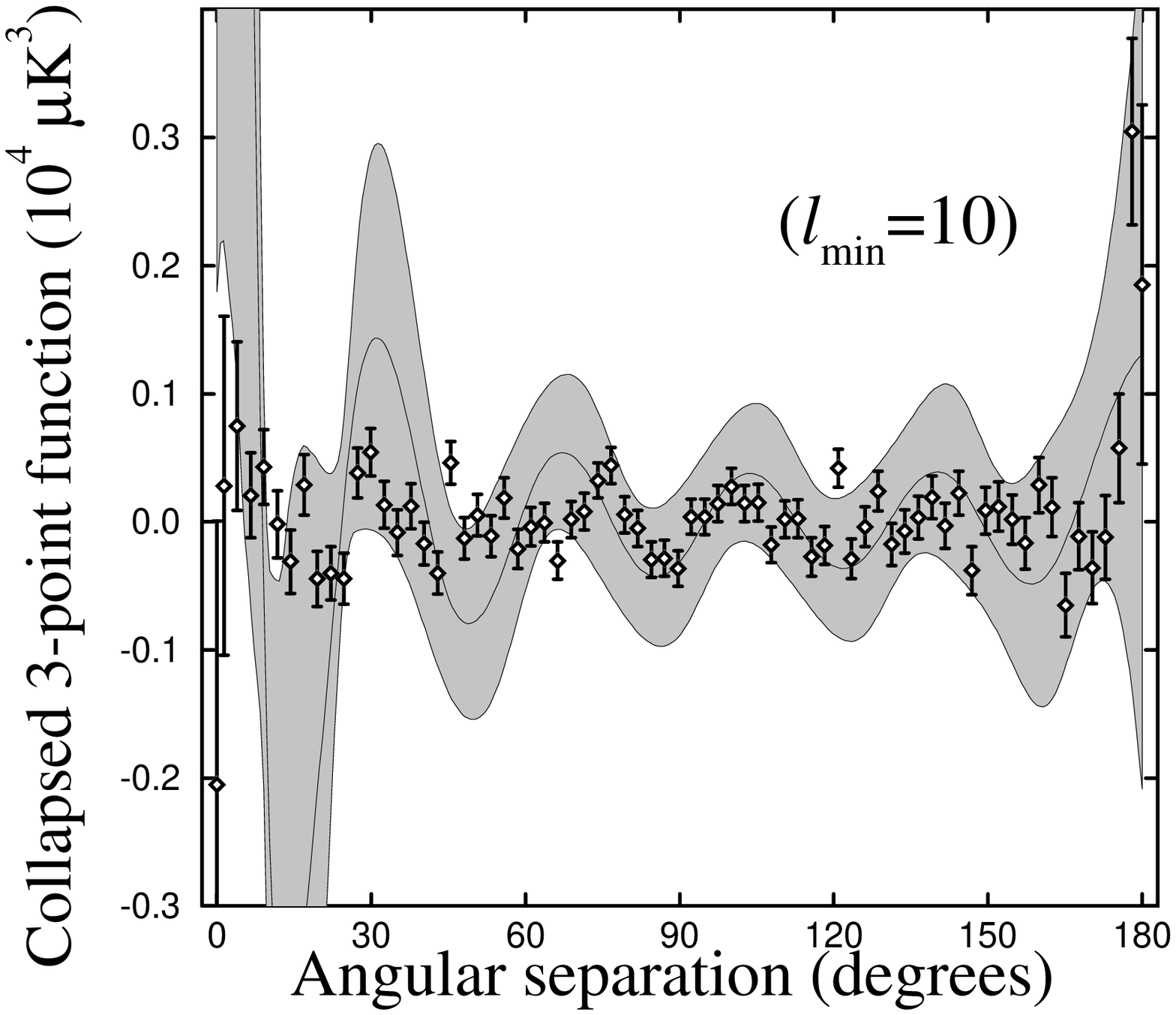} }
{\hskip -0.3cm}
{\epsfxsize = 8.5cm \epsffile{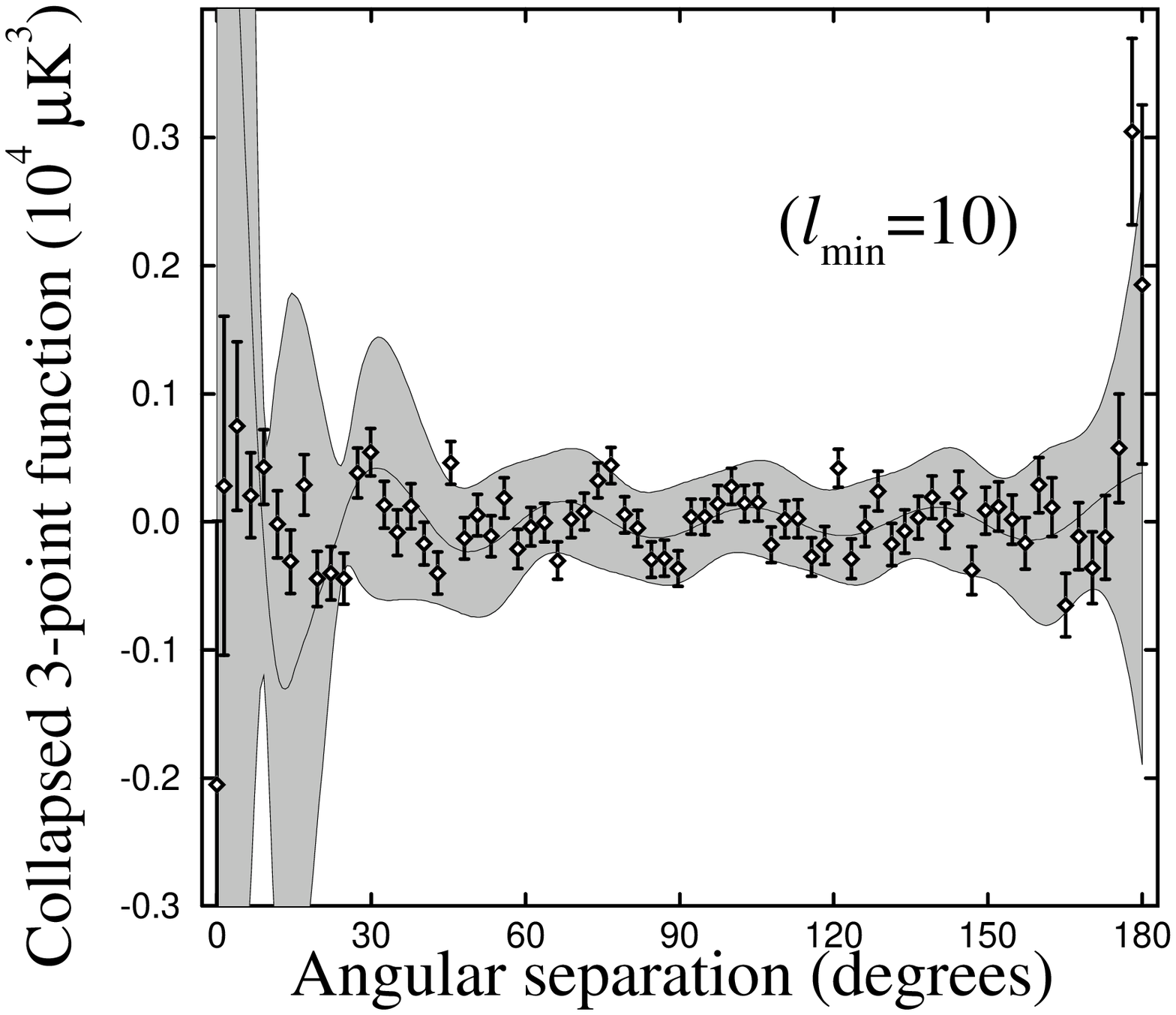} } }
\parbox{16cm}{\caption[figure3]{\small
\baselineskip 12pt
Combined four--year {\sl COBE}--DMR data and collapsed three--point
correlation function predicted by the analytical texture model (as in
previous figures). Left panel: for a somewhat exaggerated value of the
asymmetry parameter $x=0.4$. Right panel: for the value $x \sim 0.1$
suggested by texture simulations.}}
\end{figure}
From these figures one may see qualitatively by eye that 
(for some ranges of the angular separation better than for others, 
of course) the data seems to follow `approximately' 
the trend of the texture curves. 
While many of the data points
fell outside the Gaussian band (Figure 1), most of
them are now inside the grey band in Figure 3 (left panel). 
Moreover, while we vary the $x$ values from 0.4 down to 0.1
(the actual value suggested by texture simulations) we see that more 
and more points (with error bars) get inside (or touch) the grey band. 
Can this be just by chance? Or is it there something worth of further 
study? 
In order to answer these questions one ought to quantify more the 
analysis by using a $\chi^2$ statistics for the model and data, 
and comparing it to the Gaussian case, e.g.$^{11)}$ 
\begin{equation}
\chi^2 = \sum_{\alpha \beta} 
(D_\alpha - \bigl\langle C_3 \bigr\rangle_\alpha)
({\rm M}^{-1})_{\alpha \beta}
(D_\beta - \bigl\langle C_3 \bigr\rangle_\beta) ,
\end{equation}
with $D_\alpha$ the {\sl COBE}--DMR three--point function and 
${\rm M}$ is the covariance matrix of the analytical model.
It might be that the data picks out a preferred value for the
asymmetry parameter $x$.
Work in this direction is currently under way.

\newpage


\noindent
{\bf Acknowledgments:} 
I thank Gary Hinshaw and the {\sl COBE} team for providing the 4-yr
data, and especially Gary for useful correspondence.
I also thank Silvia Mollerach for her collaboration on the work
described herein, Ruth Durrer, Andrew Liddle and Neil Turok
for useful conversations during the workshop, 
and the organizers for making this such a stimulating meeting.
I acknowledge partial funding from The British Council/Fundaci\'on 
Antorchas. 

\frenchspacing
\section*{References}

\vspace*{12pt}

{\baselineskip 16pt
\noindent
1.~W. Hu, N. Sugiyama and J. Silk, astro-ph/9604166.\\
2.~A. Albrecht, D. Coulson, P. Ferreira and J. Magueijo,
   Phys. Rev. Lett. {\bf 76}, 1413 (1996); see also Hobson's
   contribution to these proceedings.\\
3.~R. G. Crittenden and N. Turok, Phys. Rev. Lett. {\bf 75}, 2642
   (1995); R. Durrer, A. Gangui and M. Sakellariadou,
   Phys. Rev. Lett. {\bf 76}, 579 (1996).\\
4.~N. Turok, astro-ph/9604172.\\
5.~J. Magueijo,  Phys. Rev. D {\bf 52}, 689 (1995).\\
6.~J. Borrill, E. Copeland, A. Liddle, A. Stebbins and
   S. Veeraraghavan,  Phys. Rev. D. {\bf 50}, 2469 (1994).\\
7.~A. Gangui, F. Lucchin, S. Matarrese and S. Mollerach,
   Astrophys. J. {\bf 430}, 447 (1994).\\
8.~E. W. Wright, C. L. Bennett, K. M. G\'orski, G. Hinshaw and
   G. F. Smoot, astro-ph/9601059.\\
9.~G. Hinshaw, A. J. Banday, C. L. Bennett, K. M. G\'orski
   and A. Kogut, Astrophys. J. {\bf 446}, L67 (1995).\\
10.~A. Gangui and S. Mollerach, astro-ph/9601069.\\
11.~A. Kogut, A. J. Banday, C. L. Bennett, K. G\'orski, G. Hinshaw, 
    G. F. Smoot and E. L. Wright, astro-ph/9601062.
}

\end{document}